\documentclass[12pt]{article}

\usepackage{amssymb}
\usepackage{amsbsy}
\usepackage{amsfonts}
\usepackage{graphicx}

\begin{document}
\addtolength{\topmargin}{-15mm}
\addtolength{\oddsidemargin}{-10mm}
\addtolength{\evensidemargin}{-10mm}

\pagestyle{empty}

\begin{center}

{\Large\bf FERMI SURFACES
OF CRYSTALS IN A HIGH MAGNETIC FIELD}\\[5mm]

{J.~BR\"UNING}

{\it Mathematisch-Naturwisenschaftliche Fakult\"at II der
Humboldt-Universit\"at zu Berlin,\\
Rudower Chaussee 25, Berlin 12489, Germany}\\[3mm]

{V.~V.~DEMIDOV}

{\it Laboratory of Mathematical Physics, Mordovian State
University,\\
Bolshevistskaya 68, Saransk, Mordoviya 430000, Russia}\\[3mm]

{V.~A.~GEYLER}

{\it Laboratory of Mathematical Physics, Mordovian State University\\
geyler@mathematik.hu-berlin.de}\\[4mm]
\end{center}

\begin{abstract}
\noindent A method of building and investigation of the Fermi surfaces
for three-dimensional crystals subjected to a uniform magnetic field is presented.
The Hamiltonian of a charged particle in the crystal is treated in the
framework of the zero-range potential theory.
The dispersion relation for the Hamiltonian is obtained in an
explicit form.\\[1mm]

\end{abstract}

\section{Introduction}

The shape of the Fermi surface determines the kinetic and  equilibrium properties
of the electron gas in the crystal matter \cite{Crackn} as well as the dynamics of a
single electron in crystal \cite{Novikov}. The shape of this
surface can be reconstructed drastically by a high magnetic
field \cite{Caulfield,House,Harr}. However,
a uniform magnetic field  changes totally the translation properties
of an electron in the crystal lattice: the appearance of a new
length scale (the magnetic length) leads to the famous
phenomena related to the
"commensurability--incommensurability" transitions.
In particular, a fractal structure arises in the diagram
describing the dependence of the two-dimansional electron spectrum on the
magnetic flux (Azbel'--Hofstadter butterfly \cite{Azbel',Hofstadter}).
It was shown recently that this fractal structure is an inherent
characteristic not only of two-dimensional spectral diagrams but also of
the "flux--energy" and "angle--energy" diagrams for a
three-dimensional magneto-Bloch electron \cite{KAK,BDG}.

The translation symmetry of the Bloch electron in a uniform magnetic field is
defined by the magnetic translation group \cite{Zak}, which has more complicated
structure in comparison with the translation group without the field. In particular, the
ordinary quasi-momentum is no longer a conserved quantity, therefore, at high magnetic
fields a modification of the definition of the Fermi surface is required. Such
a modification is given by V.~Ya.~Demikhovskii with coworkers \cite{Demi1,Demi2}. In
papers \cite{Demi1,Demi2},
the construction and investigation of the Fermi surfaces in the magnetic
Brillouin zone have been performed by means of the tight-binding approximation. In
particular, it was shown that the change of the topology of the Fermi surface under
the influence of the magnetic field can cause the metal--semiconductor transition.

The procedure of obtaining the tight-binding
Schr\"odinger operator (also called the Harper operator)
leads to a rather rough approximation of the initial
Hamiltonian; for example,
in the 2D case it is known that the "flux--energy"
diagram for the original periodic Landau operator differs
essentially  from the classical Hofstadter butterfly \cite{PK,GA}.

In this connection, it is interesting to build and investigate the
Fermi surfaces in the magnetic Brillouin zone for the periodic
Landau operator without additional approximations. Here we consider
the case of the Landau operator perturbed by a lattice $\Lambda$ of
short-range scatterers. To get an explicitly solvable model we deal
here with a limiting case of the short-range potential, namely,
with the zero-range potential \cite{BZP,DO,Albeverio}, also called
"point potential". We stress, that the method of zero-range
perturbations for obtaining the Fermi surfaces in 3D crystals
without magnetic fields was successfully used earlier \cite{Hoeg}.
On the other hand, the spectral properties of the three-dimensional
Landau operator perturbed by a periodic point potential was studied
in papers \cite{BDG,AAG,GD}. In particular, in the paper \cite{GD}, a
version of the magneto-Bloch analysis was proposed which makes it
possible to get an explicit form of the dispersion relation for the
problem we consider here.

\section{The model Hamiltonian and dispersion relation}

We start with the quantum-mechanical Hamiltonian $H_0$ of a single electron subject to
a uniform magnetic field $\mathbf{B}$
\begin{equation}
               \label{1}
H_0=\frac{1}{2\mu}\left({\bf p}-\frac{e}{c}{\bf A}({\bf r})\right)^2\,,
\end{equation}
where $e$ and $\mu$ are the charge and mass of the electron, respectively, and ${\bf
A}({\bf r})$ is the vector potential of the field ${\bf B}$. We choose the symmetric
gauge: ${\bf A}({\bf r})={\bf B}\times{\bf r}/2$; in this case the form of $H_0$ in
any Cartesian coordinate system with the $z$-axis directed along the field ${\bf B}$
is independent of the choice of the other coordinate axes.

We use the standard notations $\omega=|eB|/c\mu$ (the cyclotron frequency),
$l_M=(\hbar/\mu\omega)^{1/2}$ (the magnetic length), and $\phi_0=2\pi\hbar c/|e|$ (the
magnetic flux quantum). Throughout the paper we denote by ${\bf b}$, ${\bf b}=-{\bf
B}/\phi_0$, the vector of the magnetic flux density: if ${\bf a}_1$ and ${\bf a}_2$
are nonzero vectors, then $|{\bf b}({\bf a}_1\times{\bf a}_2)|$ is the total number of
the magnetic flux quanta through the parallelogram spanned by ${\bf a}_1$ and ${\bf
a}_2$. Note that $|{\bf b}|= (2\pi l^2_M)^{-1}$.

All important physical information concerning the Hamiltonian $H_0$
is contained in the Green function $G_0$ at energy $E$
defined as $G_0({\bf r},{\bf r}';E)=\langle{\bf r}|(E-H_0)^{-1}|{\bf r}'\rangle$. It
is known (see, e.g., Ref.~\cite{GM87,Gou}) that this function can be represented in
the following forms:
\begin{equation}
           \label{Green}
G_0({\bf r},{\bf r}';E)=\Phi({\bf r},{\bf r}')F_1({\bf r}-{\bf r}';E)= \Phi({\bf
r},{\bf r}')F_2({\bf r}-{\bf r}';E)\,,
\end{equation}
where
\begin{equation}
               \label{19}
\Phi({\bf r},{\bf r}')=\frac{\mu}{2^{3/2}\pi\hbar^2l_M} \exp\left[-\pi i{\bf b}({\bf
r}\times{\bf r}')- \frac{({\bf r}_\bot-{\bf r}_\bot')^2}{4l_M^2}\right]\,,
\end{equation}
\begin{eqnarray}
               \label{20}
F_1({\bf r};E)=\sum_{l=0}^\infty\frac{\exp\left[-\sqrt{2(l+1/2-E/\hbar\omega)} |{\bf
r}_{||}|/l_M\right]} {\sqrt{l+1/2-E/\hbar\omega}}\,L_l({\bf r}_\bot^2/2l_M^2)\,,
\end{eqnarray}
\begin{equation}
               \label{21}
F_2({\bf r};E)=\int\limits_0^\infty\frac{\exp\left[- \left({\bf r}_\bot^2/(e^t-1)+{\bf
r}_{||}^2/t\right)/2l_M^2\right]}
{\left(1-e^{-t}\right)\exp\left[(1/2-E/\hbar\omega)t\right]} \frac{dt}{\sqrt{\pi
t}}\,.
\end{equation}
Here ${\bf r}_{||}$ is the projection of ${\bf r}$ on the direction of the field ${\bf
B}$ and ${\bf r}_{\bot}={\bf r}-{\bf r}_{||}$; the function $L_l(x)$ is the $l$-th
Laguerre polynomial.

It is convenient to decompose $G_0$ into a sum of singular and regular
(with respect to the limit ${\bf r}\to{\bf r}'$) parts :
\begin{equation}
               \label{2}
G_0({\bf r},{\bf r}';E)=\frac{\mu}{2\pi\hbar^2} \frac{\exp\left[-i\pi{\bf
b}({\bf r}\times{\bf r}')\right]} {|{\bf r}-{\bf r}'|}+
G_0^{\rm reg}({\bf r},{\bf r}';E)\,.
\end{equation}
Note that $G_0^{reg}({\bf r},{\bf r};E)$ is well defined at ${\bf
r}={\bf r}'$ and
\begin{equation}
                 \label{sGr}
G_0^{reg}({\bf r},{\bf
r};E)=\frac{\mu}{2^{3/2}\pi\hbar^2l_M}\,\zeta\left({1\over2},{1\over2}
-{E\over\hbar\omega}\right)\,,
\end{equation}
were ${\zeta}(s,v)$ is the generalized Riemann zeta-function (in other words, the
Hurwitz zeta-function) \cite{BE}.

Now we consider a crystalline lattice $\Gamma$ with nodes $\boldsymbol{\gamma}$; the
Bravais
lattice of the crystal we denote by $\Lambda$. Choose a basis ${\bf a}_1$, ${\bf
a}_2$, ${\bf a}_3$ of $\Lambda$, then a basic cell $C_\Gamma$ of $\Gamma$ (the
Wigner--Seitz cell) is fixed: $C_\Gamma=\{t_1{\bf a}_1+t_2{\bf a}_2+t_3{\bf a}_3:\ 0\le t_i< 1\}$.
If ${\rm K}$ is the set of nodes of $\Gamma$ containing in $C_\Gamma$,
then $\Gamma={\rm K}+\Lambda$; this means that each vector
$\boldsymbol{\gamma}\in\Gamma$ can be represented in the form
$\boldsymbol{\gamma}=\boldsymbol{\kappa}+\boldsymbol{\lambda}$, where $\boldsymbol{\kappa}\in{\rm K}$,
$\boldsymbol{\lambda}\in\Lambda$ (note that such a representation is unique). Without loss of
generality we will suppose that ${\bf 0}\in {\rm K}$.
We choose the potential $V({\bf r})$ of the crystalline lattice $\Gamma$ in the form
\begin{equation}
                     \label{poten1}
V({\bf r})= \sum\limits_{\boldsymbol{\kappa}\in\,{\rm
K}}\sum\limits_{\boldsymbol{\lambda}\in\Lambda}V_{\boldsymbol{\kappa}}({\bf
r}-\boldsymbol{\lambda})\,,
\end{equation}
where $V_{\boldsymbol{\kappa}}({\bf r})$ is the confinement potential near the node
$\boldsymbol{\kappa}$. To obtain an explicitly solvable model, we pass to the zero-range
limit for $V_{\boldsymbol{\kappa}}$. This means that we consider $V_{\boldsymbol{\kappa}}$ as
the limit of potentials of the form $c_{\boldsymbol{\kappa}}W({\bf r}-\boldsymbol{\kappa})$, where
$W({\bf r})\sim 0$ outside a small sphere of radius $R$ centered at zero, the coupling
constant $c_{\boldsymbol{\kappa}}$ is of order $R$, and $\int W({\bf r})\,d{\bf r} =1$. At the
limit $R\to 0$ the potential $V_{\boldsymbol{\kappa}}$ is characterized by one parameter only,
namely, by the scattering length $\rho_\kappa$, which is related to the binding energy
of the ground state for $V_{\boldsymbol{\kappa}}$ by
$$
E_{\boldsymbol{\kappa}}=-\frac{\hbar^2}{2\mu\rho_{\boldsymbol{\kappa}}^2}
$$
(see Ref.~\cite{BZP}). What is important,
at the zero-range limit the Green function $G({\bf r},{\bf
r}';E)$ of $H$ has the following explicit expression in terms of the Green function
$G_0({\bf r},{\bf r}';E)$ of $H_0$ (see Ref.~\cite{AAG,GM87}):
\begin{eqnarray}
                \label{5}
G({\bf r},{\bf r}';E)=G_0({\bf r},{\bf r}';E)
-\sum_{\boldsymbol{\gamma},\boldsymbol{\gamma}'\in\Gamma} G_0({\bf
r},{\boldsymbol{\gamma}};E)\left(S^{-1}(E)\right)_{\boldsymbol{\gamma},\boldsymbol{\gamma}\,{}'}
G_0(\boldsymbol{\gamma}\,{}',{\bf r}';E),
\end{eqnarray}
where $S^{-1}(E)$ is the matrix inverse to the infinite matrix $S(E)$ with elements
\begin{eqnarray}
               \label{6}
S(\boldsymbol{\gamma},\boldsymbol{\gamma}\,{}';E)=\left[G_0^{reg}(\boldsymbol{\gamma},\boldsymbol{\gamma};E)-
\frac{\mu}{2\pi\hbar^2\rho_{\boldsymbol{\gamma}}}
\right]\delta_{\boldsymbol{\gamma},\boldsymbol{\gamma}\,{}'}+
(1-\delta_{\boldsymbol{\gamma},\boldsymbol{\gamma}\,{}'})G_0(\boldsymbol{\gamma},\boldsymbol{\gamma}\,{}';E)\,.
\end{eqnarray}

The Hamiltonian $H=H_0+V$ in some sense is a three-dimensional analogue of the
one-dimensional Kronig--Penney Hamiltonian. The spectral properties of $H$ can be
extracted from the properties of invertibility of the infinite matrix $S(E)$. Namely,
at least for $E<\hbar\omega/2$ the energy $E$ belongs to the spectrum of $H$ if the
matrix $S(E)$ is not invertible. Under so-called "rationality
condition" \cite{Zak}, we can reduce the problem to invert the infinite
matrix $S(E)$ to a problem of finite-dimensional algebra.
According to \cite{Zak}, we call the field ${\bf B}$ {\it rational}
with respect to $\Lambda$ if for some basis
${\bf a}_1$, ${\bf a}_2$, ${\bf a}_3$ of $\Lambda$, all the
numbers ${\bf b}({\bf a}_j\times{\bf a}_k)$ are rational. It is
clear that every ${\bf B}$ obeys this condition after an
appropriate infinitely small change of direction, therefore, the
rationality condition imposes no essential restrictions on ${\bf B
}$. If ${\bf B}$ is rational with respect to
$\Lambda$, then a basis ${\mathbf a}_j$ of $\Lambda$ can be chosen in such a
way that $\eta\equiv\mathbf{b}({\mathbf a}_1\times {\mathbf a}_2)>0$ and
$\mathbf{b}({\mathbf a}_2\times {\mathbf a}_3)=\mathbf{b}({\mathbf a}_1\times {\mathbf
a}_3)=0$. Denote $\eta=N/M$, where $N$ and $M$ are coprime positive integers;
in our approach the crucial role is played by the following
$(MK)\times(MK)$ matrix $\widetilde S$, where $K$ is the number of nodes in ${\rm K}$:
$$
\widetilde S_{q,q'}({\bf p},E)=\exp[-\pi i m' {\bf b}(\boldsymbol{\kappa}'\times {\bf
a}_2)]\times\\
$$
$$
\sum\limits_{\lambda_1, \lambda_2, \lambda_3=-\infty}^{\infty} S(\lambda_1{\bf a}_1
+(\lambda_2M + m){\bf a}_2 + \boldsymbol{\kappa},\,m'{\bf a}_2
+\boldsymbol{\kappa}',\,\lambda_3{\bf a}_3;E)\times\\
$$
\begin{eqnarray}
        \label{KPR}
\exp\left[\pi i(\lambda_1{\bf a}_1 + (\lambda_2M + m){\bf a}_2)({\bf
b}\times\boldsymbol{\kappa}-\eta\lambda_1(M\lambda_2 + m )-2\boldsymbol{\lambda}\cdot{\bf p}
)\right]\,.
\end{eqnarray}
Here $q$ denotes the pair $(m,\boldsymbol{\kappa})$ with $\boldsymbol{\kappa}\in{\rm K}$,
$m=0,\ldots,M-1$ and ${\bf p}$ is the magnetic quasimomentum.
The magnetic Brillouin zone is defined by the inequalities $0\le
p_1\le M^{-1}$, $0\le p_2\le 1$, $0\le p_3\le 1$ and, in this zone, the
dispersion relation for $H$ reads
\begin{equation}
      \label{KP}
  {\rm det}\,\widetilde S({\bf p},E)=0\,.
\end{equation}
For a fixed $\mathbf{p}$, equation (\ref{KP}) has  infinitely many solutions
$E_s(\mathbf{p})$, $s=0,1,\ldots$;
each $E_s(\mathbf{p})$ is an $M$-fold degenerate eigenvalue of $H$.
Dispersion laws $E=E_s(\mathbf{p})$ are continuous functions of
$\mathbf{p}$; the set of all values of $E_s(\mathbf{p})$ is
an interval $J_s$. In a natural way, the intervals $J_s$ form clusters
called {\it magnetic bands} or {\it Landau bands}, each magnetic band
contain $KM$ adjacent intervals $J_s$ ({\it subbands}).
  For the Fermi energy $E_F$, the equation
\begin{equation}
         \label{FE}
E_s(\mathbf{p})=E_F
\end{equation}
determines a periodic surface in the quasimomentum space (the {\it Fermi surface}).

\section{Results of numerical analysis}

Using Eqs.~(\ref{KPR}) --- (\ref{FE}) we analyze numerically the shape of the
Fermi surfaces
in a crystal with the simple-cubic lattice for various directions of the field ${\bf
B}$. The lattice constant $a=7.5$~nm is chosen relevant to the geometric parameters
for the 3D regimented quantum dot superlattice considered recently in Ref.~\cite{Lazar}.
As to the scattering length, we put $\rho\sim 1$~nm, this corresponds to the binding
energy $E\sim30$~meV.

Let the magnetic field ${\bf B}$ is directed along the edge ${\bf a}_3$ of the cubic
elementary cell and $\eta=1$; this value of $\eta$ corresponds to the field
strength $\thicksim70$~T (note in this connection that it was
reported on the possibility to achieve the magnetic field strength
28 MG \cite{Boyko}). In the considered case, the lowest magnetic
band of the spectrum has the form $[-1.21\varepsilon_0;\
-0.68\varepsilon_0]$, where $\varepsilon_0=\hbar\omega/2$.
The Fermi surfaces for various energy values from this band are
depicted in Fig.~1. The warped ellipsoids in the panels (a) and (d)
are the electron and hole surfaces respectively; in this case {\it
there exist only closed trajectories of the charged particle in
the $p$-space.} Other topologies of the Fermi surface are shown in
panels (b) and (c); according to the definition from \cite{Novikov}
these surfaces have the topology of range 2 and 3 respectively.

Surprisingly, an inclination of ${\bf B}$ to the edge of the cubic
cell causes a profound reconstruction of the Fermi surfaces.
As an example we consider ${\bf B}=B(0, 3/5, 4/5)$. In the new basis ${\bf a}'_1={\bf
a}_1$, ${\bf a}'_2={\bf a}_2+{\bf a}_3$, ${\bf a}'_3=3{\bf a}_2+4{\bf a}_3$,
we have ${\bf a}'_3||{\bf B}$ and $\eta=1/5$. Therefore, the
magnetic Brilluoin zone is defined by the inequalities $0\le
p_1\le 1/5$, $0\le p_2,p_3\le 1$, and the lowest magnetic band
splits into 5 subbands.
Fig.~2 shows variation of the Fermi surface as the Fermi energy $E_F$ varies
in the first subband $[-1.11\,\varepsilon_0;\,-0.98\,\varepsilon_0]$.
The electron and hole Fermi surfaces
have the shapes of warped cylinders (panels (a) and (d)
respectively) or highly twisted sheets (panels (b) and (c)).
It is particularly remarkable that  in this case
the surfaces {\it are open in the $p_1$-direction}.
In going from the first subband  of the lowest magnetic band to the
higher subbands, the shape of the Fermi surface changes; examples
of this change are shown on Fig.~3, where the Fermi energy  $E_F$ lies in
the second and fourth subbands respectively.

In the examples above, the filed ${\bf B}$ is parallel to a face of the
cubic cell. We have built the Fermi surface for the field of the
form ${\bf B}=B(12/25,9/25,4/25)$, where $\eta=1/25$ but by the
technical reasons do not give the corresponding figure here. In
this case, the Fermi surface is divided into great numbers of
connected pieces having the shape of winding sheets. Such a
complex structure of the Fermi surface reflects the fractal
structure of the spectrum of the periodic Landau
operator \cite{BDG}.

\section*{Acknowledgments}

We are thankful to S.~Albeverio, M.~V.~Budantsev, V.~Ya.~Demikhovskii,
and P.~Exner for valuable discussions and useful remarks.
This work was supported by Grants of INTAS 00-257, DFG 436 RUS
113/572/0-2, and RFBR 02-01-00804.


\begin{thebibliography}{0}
\bibitem{Crackn} A. P. Cracknell and K. C. Wong, {\it The Femi surface}
               (Clarendon Press, Oxford, 1973).
\bibitem{Novikov} S.~P.~Novikov, A.~Ya.~Maltsev. {\it Uspekhi Fiz. Nauk} {\bf 168}, 249 (1998).
\bibitem{Caulfield} J.~Caulfield, J.~Singleton, P.~T.~J.~Hendriks, J.~A.~A.~J.~Perenboom,
                    F.~L.~Pratt, M.~Doporto, W.~Hayes, M.~Kurmoo, and P.~Day, {\it
                    J. Phys.: Condens. Matter}, {\bf 6}, L155 (1994).
\bibitem{House} A.~A.~House, N.~Harrison, S.~J.~Blundell, I.~Deckers, J.~Singleton, F.~Herlach,
                 W.~Hayes, J.~A.~A.~J.~Perenboom, M.~Kurmoo, and P.~Day, {\it Phys.
                 Rev. B}, {\bf 53}, 9127 (1996).
\bibitem{Harr} N.~Harrison, D.~W.~Hall, R.~G.~Goodrich, J.~J.~Vuillemin and Z.~Fisk,
              {\it Phys. Rev. Lett}. {\bf 81}, 870 (1998).
\bibitem{Azbel'} M.~Za.~Azbel', {\it Zhurn. Exper. i Theor. Fiz.},
                 {\bf 46}, 929 (1964) [in Russian].
\bibitem{Hofstadter} D.~R.~Hofstadter, {\it Phys. Rev. B} {\bf 14}, 2239 (1976).
\bibitem{KAK} M.~Koshino, H.~Aoki, K.~Kuroki, S.~Kagoshima,
           and T.~Osada, {\it Phys. Rev. Lett.} {\bf 86}, 1062 (2001);
            M.~Koshino, H.~Aoki, T.~Osada, K.~Kuroki, and S.~Kagoshima,
           {\it Phys. Rev. B} {\bf 65}, 045310 (2002); M.~Koshino and
            H.~Aoki, {\it Phys. Rev. B} {\bf 67}, 195336 (2003).
\bibitem{BDG} J.~Br\"uning, V.~V.~Demidov, and V.~A.~Geyler,
           {\it Spectral diagrams of Hofstadter type for the Bloch electrons in three
           dimensions.}  \verb"cond-mat/0306391".
\bibitem{Zak}  J.~Zak, {\it Phys. Rev. A} {\bf 134}, 1602 (1964).
\bibitem{Demi1} V.~Ya.~Demikhovskii, A.~A.~Perov, and D.~V.~Khomitsky,
               {\it Phys. Lett. A}, {\bf 267}, 408 (2000).
\bibitem{Demi2} V.~Ya.~Demikhovskii and D.~V.~Khomitsky, {\it Zhurn. Exper. i Theor.
                Fiz.}, {\bf 120}, 191 (2001) [in Russian].
\bibitem{PK}   G.~Petschel and T.Geisel, {\it Phys. Rev. Lett.}, {\bf
               71}, 239 (1993); O.~K\"uhn, V.~Fessatidis,
               H.~L.~Cui, P.~E.~Selbmann, and N.~J.~M.~Horing, {\it
               Phys. Rev. B} {\bf 47}, 13019 (1993).
\bibitem{GA}   S.~A.~Gredeskul, M.~Zusman, Y.~Avishai, and
               M.~Ya.~Azbel, {\it Phys. Rep.}, {\bf 288} (1997),
               223;
               V.~A.~Geyler, I.~Yu.~Popov, A.~A.~Ovechkina,
               and A.~V.~Popov, {\it Chaos, Solitons, and Fractals}
               {\bf 11}, 281 (2000).
\bibitem{BZP} A.~I.~Baz', Ya.~B.~Zeldovich, and A.~M.~Perelomov,
              {\it Scattering, reactions and decay in
              nonrelativistic quantum mechanics} (Moscow, Nauka, 1971).
\bibitem{DO} Yu.~N.~Demkov and V.~N.~Ostrovskiy, {\it Zero-range
                potentials and their applications in atomic physics}
                (Plenum Press, New York, 1988).
\bibitem{Albeverio} S.~Albeverio, F.~Gesztesy, R.~H{\o}egh-Kron and H.~Holden, {\it Solvable
                    models in quantum mechanics} (Springer-Verlag, Berlin, 1988).
\bibitem{Hoeg} R.~H{\o}egh-Kron, H.~Holden, S.~Johannesen and T.~Wentzel-Larsen,
               {\it J. Math. Phys.} {\bf 27}, 385 (1986).
\bibitem{AAG}  Y.~Avishai M.~Ya.~Azbel, S.~A.~Gredeskul,
                {\it Phys. Rev. B}, {\bf 48}, 17280 (1993).
\bibitem{GD}  V.~A.~Geyler and V.~V.~Demidov, {\it Theor. Math. Phys.} {\bf 103}, 561 (1995).
\bibitem{GM87}  V.~A.~Geyler and V.~A.~Margulis, {\it Theor. Math. Phys.} {\bf 70}, 133 (1987).

\bibitem{Gou}   G.~Gountarulis, {\it Phys. Lett. A} {\bf 36}, 132 (1972).

\bibitem{BE} H.~Bateman and A.~Erd{\'e}lyi, {\it Higher transcendental
                functions, V.~I} (McGraw-Hill, New York, 1953).
\bibitem{Lazar} O.~L.~Lazarenkova and A.~A.~Balandin, {\it Phys. Rev. B}
               {\bf 66}, 254319 (2002).
\bibitem{Boyko} B.~A.~Boyko, A.~I.~Boyko, M.~I.~Dolotenko et al., Book of abstracts,
                {\it VIII Int. Conference on Megagauss Magnetic Field and Related
                Topics,} (Tallahassee, USA), p.~149.
\end{thebibliography}
\end{document}